\documentclass[aps,prl,twocolumn,showpacs,superscriptaddress,floatfix]{revtex4-1}

\usepackage{graphicx}

\usepackage{bm}        
\usepackage{epstopdf}

\usepackage{subfig}
\usepackage{amsmath}	
\usepackage{amssymb}    

\newcommand{\be}{\begin{eqnarray}}
 \newcommand{\ee}{\end{eqnarray}}

\begin{document}



\title{Formal analogy between the Dirac equation in its Majorana form and the discrete-velocity version of the Boltzmann kinetic equation}

\author{F. Fillion-Gourdeau}
\email{filliong@CRM.UMontreal.ca}
\affiliation{Centre de Recherches Math\'{e}matiques, Universit\'{e} de Montr\'{e}al, Montr\'{e}al, Canada, H3T 1J4}

\author{H.J. Herrmann}
\affiliation{ETH Zurich, Computational Physics for Engineering Materials,
Institute for Building Materials, Schafmattstrasse 6, HIF, CH-8093 Zurich (Switzerland)}

\author{M. Mendoza}
\affiliation{ETH Zurich, Computational Physics for Engineering Materials,
Institute for Building Materials, Schafmattstrasse 6, HIF, CH-8093 Zurich (Switzerland)}

\author{S. Palpacelli}
\email{silviapalpacelli@gmail.com}
\affiliation{Numidia srl, via G. Peroni, 130, 00131, Roma, Italy}

\author{S. Succi}
\email{succi@iac.cnr.it}
\affiliation{Istituto Applicazioni Calcolo, CNR, via dei Taurini 19, 00185, Roma, Italy}
\altaffiliation[Also at ]{Physics Department, Harvard University, Cambridge MA, 02138, USA}

%
\date{\today}

\begin{abstract}
We point out a formal analogy between 
the Dirac equation in Majorana form and the discrete-velocity
version of the Boltzmann kinetic equation.
By a systematic analysis based on the theory of operator splitting, this  
analogy is shown to turn into a concrete and efficient computational 
method, providing a unified treatment of relativistic and non-relativistic quantum mechanics. 
This might have potentially far-reaching implications for both classical and quantum
computing, because it shows that, by splitting time along the three spatial directions, 
quantum information  (Dirac-Majorana wavefunction) propagates in space-time as a 
classical statistical process (Boltzmann distribution).

\end{abstract}


\maketitle


\pagestyle{plain}


\section{Boltzmann and Dirac}

Analogies between the non-relativistic Schr\"{o}dinger equation and fluid dynamics 
have been noted since the early days of quantum mechanics. 
In particular, back in 1927, Erwin Madelung noticed that by expressing the 
wavefunction in eikonal form, i.e. $\Psi=R \;e^{iS/\hbar}$, the Schr\"{o}dinger
equation turns into the hydrodynamic equation of a compressible, inviscid
fluid, with number density $\rho=R^2$ and velocity $\vec{u}=-\nabla S/m$. 
The quantum fluid is subject to the classical potential 
$V_c(\vec{x})$, plus the quantum potential 
$V_q(\vec{x})=-\frac{\hbar^2}{2m} (\Delta R)/R$.
Although the hydrodynamic analogy is commonly regarded as purely 
formal in nature, lately, its connections with Bohm's theory
of hidden variables and De Broglie's pilot wave picture have known
of surge of interest, mostly in connection with experimental investigations
on the non-local nature of quantum physics \cite{HAROCHE}.

The quantum relativistic fluid analogy seems to 
have received comparatively less attention.
Back in 1993, it was noted that the Dirac equation can be regarded
as a special form of a {\it discrete} Boltzmann kinetic equation, in which
the particle velocities are confined to a handful of discrete values \cite{Succi1993327}.
The discrete components of the Boltzmann distribution, 
$f_i(\vec{x};t) \equiv f(\vec{x},\vec{v}=\vec{v}_i;t)$,
where the index $i$ labels the discrete velocities, are then
identified with the spinor components $\psi_i$ of the Dirac equation.
This opens up an interesting connection between classical kinetic theory
and relativistic quantum mechanics.


Mathematically, the connection is not so surprising, since
both Boltzmann and Dirac equations are hyperbolic supersets
of the Navier-Stokes and Schr\"{o}dinger equations, respectively.

The interesting point, however, is that the connection becomes much 
more direct and compelling by considering the discrete-velocity 
version of the Boltzmann equation, in relation to the Majorana form 
of the Dirac equation, in which all matrices are {\it real} \cite{Itzykson:1980rh}.  

Majorana particles have attracted significant interest in recent years,
mostly in connection with the fact that they coincide with their own
antiparticles, as beautifully discussed in a recent essay 
by F. Wilczek \cite{WILCZ}.

Here, we wish to put forward a different angle of interest of the
Majorana representation, namely the fact that not only it makes 
Boltzmann-Dirac analogy conceptually more poignant, but it also
turns it into a concrete {\it unified} computational scheme for 
the simulation of both relativistic and non-relativistic
quantum wave equations, on both classical and quantum computers.
The corresponding method is known as quantum lattice Boltzmann 
(QLB) method \cite{Succi1993327}.
The QLB is based on the identification of the discrete Boltzmann distribution
with the spinorial wavefunction $f_i(\vec{x};t) \leftrightarrow \psi_i(\vec{x};t)$. 
Even though both objects are real, they still face
a mismatch of degrees of freedom in more than one spatial dimensions,
since a spinor of order $s$ consists of $2s+1$ components, regardless
of the number of dimensions, while the discrete distribution requires (at least)
$2d$ discrete components in $d$ spatial dimensions.
Moreover, the Dirac-Majorana matrices cannot be simultaneously 
diagonalized, reflecting the basic fact that spinors 
are not ordinary vectors. 
As a result, in more than one spatial
dimensions, it is in principle not possible
to keep the particle velocity aligned with its spin.

Remarkably, both problems can be circumvented by resorting to operator splitting.
Essentially, this amounts to splitting the spinor propagation along the three
spatial dimensions into a series of three one-dimensional propagations,
each using the diagonalized form of the corresponding
Dirac-Majorana streaming matrix.    
As a result, at each propagation step the particle spin is kept 
aligned with its velocity, so that the identification
$f_i \leftrightarrow \psi_i$ continues to hold.  

In this Letter we show that this ``heuristic stratagem'' 
is backed up by a rigorous mathematical treatment, which leads to 
a {\it unified} computational approach to quantum wave mechanics.
The resulting computational scheme offers outstanding amenability
to parallel computing on electronic computers \cite{FillionGourdeau20121403} and is also suitable
to prospective quantum computing simulations \cite{feynman1982simulating,SLoyd,boghosian1998simulating}. 

To show its versatility also towards the inclusion of non-linear interactions,
as an application, we shall solve a specific form of the non-linear
Dirac equation including dynamical-symmetry breaking term, as first
proposed by Nambu and Jona-Lasinio.

{\it Discrete Boltzmann and Dirac}
  
To set up the framework, let us write down the two equations in full display.
The discrete Boltzmann equation reads as follows:
\begin{equation}
\label{BOL}
\partial_t f + v^a_i \nabla_a f_i = \Omega_{ij} (f_j-f_j^e)
\end{equation}
where $f_i=f(\vec{x},\vec{v}=\vec{v}_i;t)$ is the probability density of finding a particle
around position $\vec{x}$ at time $t$ with discrete velocity $\vec{v}_i$. 
The latin index $a=x,y,z$ runs over spatial dimensions and
Einstein summation rule is assumed.  
The left hand side represents the
particle free-streaming (in the absence of external forces, for
simplicity), while the right-hand side is the collisional step
steering the distribution function towards a local Maxwell equilibrium
$f_i^{e}$.
The (symmetric) scattering matrix $\Omega_{ij}$ encodes the 
mass-momentum-energy conservation laws underpinning fluid dynamic behavior.

The Dirac equation, in Majorana form, reads as follows
\begin{equation}
\label{DIRAC}
\partial_t \psi_i + S^a_{ij} \nabla_a \psi_j = M_{ij} \psi_j
\end{equation}
where $S^a_{ij}$ are the three Majorana streaming matrices and
$M_{ij}$ is the (anti-symmetric) mass matrix, acting upon the 
real spinor $\psi_i$, $i=1,2s+1$.  
This clearly shows a formal analogy with the Boltzmann equation: the lhs describes 
the free streaming of the spinors, while the rhs can be regarded as a simple form of
local collision between the various spinorial components.      
Note that the mass matrix has dimensions of an inverse time
scale, typically given by the Compton frequency $\omega_c = mc^2/\hbar$. In 1D, this analogy is ``exact'': by choosing a representation where the Dirac matrix is diagonal (Majorana representation), we recover Eq. \eqref{BOL}. 
In multiple dimensions however, the story is different: the connection can be 
realized only by resorting to operator splitting, whereby each step can be written 
in the form of Eq. \eqref{BOL}. This will be discussed in the following.


{\it Quantum lattice Boltzmann}

Let us consider the case of spin $s=1/2$ particles and start from a relativistic wave equation with matrices $\beta,\alpha_{a}$ in the Dirac representation. The goal here is to find the discrete time evolution of the wave function by using the formal analogy with the Boltzmann equation. In the QLB setting, this time evolution proceeds by a sequence of streaming and collisional steps, given by (we use natural units where $c = \hbar =1$):
\begin{eqnarray}
\label{eq:split1}
\partial_{t} \psi^{(x)}(t) &=& - \alpha_{x} \partial_{x}  \psi^{(x)}(t), \; \psi^{(x)}(t_{n}) = \psi(t_{n}),   \\
\label{eq:split2}
\partial_{t} \psi^{(y)}(t) &=&- \alpha_{y} \partial_{y}  \psi^{(y)}(t), \; \psi^{(y)}(t_{n}) = \psi^{(x)}(t_{n+1}),  \\
\label{eq:split3}
\partial_{t} \psi^{(z)}(t) &=& - \alpha_{z} \partial_{z}  \psi^{(z)}(t), \; \psi^{(z)}(t_{n}) = \psi^{(y)}(t_{n+1}),\\
\label{eq:split4}
\partial_{t} \psi^{(c)}(t) &=& -i\beta m  \psi^{(c)}(t), \; \psi^{(c)}(t_{n}) = \psi^{(z)}(t_{n+1}),   \\
 \mbox{and} &&  \psi(t_{n+1}) = \psi^{(c)}(t_{n+1}),
\end{eqnarray}
where the superscript labels the step of the splitting and $t_{n}= n \Delta t$ is the time after $n$ iterations. In these equations, the calculated solution at a given step provides an initial condition for the next step in the sequence. Eqs. \eqref{eq:split1} to \eqref{eq:split3} correspond to streaming while the last step in Eq. \eqref{eq:split4} is collisional. 

The streaming steps for a given coordinate $a$ proceed as follows. First, it should be noted that the matrix $\alpha_{a}$ (for $a=x,y,z$) is not diagonal and thus, the Dirac equation is not in the form of Eq. \eqref{BOL}. However, the latter can be recovered by using the unitary transformation of spinors 
$
S_{a} = \frac{1}{\sqrt{2}} (\beta + \alpha_{a}).
$
This equation allows to transform the Dirac matrices to a Majorana-like representation, where the 
matrix $\tilde{\alpha}_{a} =S_{a}^{\dagger}\alpha_{a}S_{a}= \beta$ is diagonal, with eigenvalues $\pm 1$. 
Then, by introducing the transformed spinor as $\tilde{\psi}^{(a)} = S^{-1}_{a}\psi^{(a)}$, the streaming steps can be turned into
\begin{eqnarray}
\partial_{t} \tilde{\psi}^{(a)}(t) &=& - \beta \partial_{a}  \tilde{\psi}^{(a)}(t),
\end{eqnarray}
which is clearly in the form of Eq. \eqref{BOL} without collisional term. This has a solution given by
\begin{eqnarray}
\tilde{\psi}_{1,2}^{(a)}(t_{n+1},\mathbf{x}) &=& \tilde{\psi}_{1,2}^{(a)}(t_{n},x_{a} - \Delta t), \\
\tilde{\psi}_{3,4}^{(a)}(t_{n+1},\mathbf{x}) &=& \tilde{\psi}_{3,4}^{(a)}(t_{n},x_{a} + \Delta t).
\end{eqnarray} 
where $x_{a}+ v^a_i = x_{a} \mp \Delta t$, $i=-1,1$, is the lattice neighbor
pointed by the discrete speed $v^a_i = \mp c$.
This corresponds to an {\it exact} integration of the streaming
operator along the characteristics $\Delta x_{a}=\pm c \Delta t$ (light-cones), which
is typical of the Lattice Boltzmann (LB) method.

The collision step can also be integrated exactly by using the solution
\begin{eqnarray}
\psi^{(c)}(t_{n+1}) = e^{-i \beta m \Delta t} \psi^{(c)}(t_{n}) \equiv C \psi^{(c)}(t_{n}).
\end{eqnarray}
It is then possible to write $C=e^{-M \Delta t}$ explicitly as a 4$\times$4 matrix by using properties of Dirac matrices\footnote{It is given by $C = \cos (m \Delta t) - i \beta \sin(m \Delta t)$}. 

It is readily shown that the above discrete system is unitary for any value of the time-step
$\Delta t$. Moreover, it looks like a classical motion of two
discrete walkers, hopping by one lattice unit along every coordinate at each
time-step and colliding according to the scattering matrix $M = i\beta m$. More complex interactions can be treated in a similar way by including the interaction terms into the scattering matrix. As long as the matrix is local, it is not necessary to diagonalize $S$ and $M$ simultaneously
and due to the operator splitting, the simplicity of the LB formalism is not compromised. For instance, for the coupling to an electromagnetic 
field, the scattering matrix is given by $M=i\beta m - i e \alpha_{a}A_{a}(x,t) + ieV(x,t)$ where $(A_{a},V)$ is the electromagnetic potential. 

Symbolically, the 3D evolution of the Dirac spinor reads like a sequence of three
one-dimensional stream steps and one collisional step:
\begin{equation}
\label{eq:QLB_exp}
\psi (t_{n+1}, \mathbf{x}) = C(S_{z}P_{z}S_{z}^{-1})(S_{y}P_{y}S_{y}^{-1})(S_{x}P_{x}S_{x}^{-1}) \psi (t_{n}, \mathbf{x}) 
\end{equation}
where $P_a=e^{-\Delta t \beta \partial_{a}}$ is a translation operator along the direction $a$. The latter shifts the ``1,2'' and ``3,4'' spinor components by $\mp \Delta t$, respectively.

Of course, this procedure is not exact: as shown in the following, it corresponds to an operator splitting method where the streaming and
collision matrices do not commute. However, each step of the splitting -is- exact and thus, the only source of error comes from the splitting which scales like $O(\Delta t^{2})$
(second order accuracy). We refer the reader to \cite{Lorin2011190} for the numerical analysis of the scheme. Other schemes where the error scales like $O(\Delta t^{3})$ can also be obtained \cite{FillionGourdeau20121403,Lorin2011190}. Most importantly, it does
not spoil the unitarity of the scheme for any value of the timestep: this is required to conserve the probability density ($L^{2}$ norm). Full details of the algorithm can be found in \cite{FillionGourdeau20121403} and slightly different versions are in \cite{Succi1993327,DELLAR}.



{\it The general operator-splitting framework}

The QLB was derived on heuristic grounds, based on a intuitive analogy
between a genuinely {\it quantum} variable, the particle spin, and 
a {\it discrete} one, the particle momentum in the lattice formulation
of the Boltzmann equation.
Since quantization is a physical concept while discretization is a numerical 
one, it might be argued that the analogy is somewhat artificial, hence perhaps 
coincidental and of limited applicability.

In the sequel, we shall show that this is not the case: 
QLB can be shown to fall within the {\it general} theory 
of operator splitting, as applied to the Dirac equation.

This might have potentially deep implications for both classical and quantum
computing, because it implies that, by splitting time along the three spatial directions, and 
augmenting the stream-collide dynamics with proper global rotations, quantum information 
(the Dirac wavefunction) propagates in space-time as a classical statistical process (Boltzmann distribution).
It would be of great interest to explore whether such insight could be used to
simulate the Dirac equation on trapped-ion analogue computers based on the QLB dynamics
\cite{gerritsma2010quantum}.

The starting point of the general operator splitting theory is the formal solution of the Dirac equation given by
\begin{eqnarray}
\psi(t_{n+1}) &=& T \exp \left[ -i \int_{t_{n}}^{t_{n+1}} H(t) dt \right] \psi(t_{n}), \\
\label{eq:suzuki_time}
&=& e^{-i\Delta t (H(t_{n}) + \mathcal{T})} \psi(t_{n})
\end{eqnarray}
where $H(t)$ is the Dirac Hamiltonian, $T$ is the time-ordering operator and $\mathcal{T} = i \overleftarrow{\partial_{t_{n}}}$ is the ``left'' time-shifting operator. The second form of the solution was obtained in \cite{suzuki1993general} and constitutes a great starting point for deriving approximation schemes. Then, the operator splitting method consists in decomposing the Hamiltonian as $H(t) = \sum_{j=1}^{N}H_{j}(t)$ and to approximate the evolution operator in Eq. \eqref{eq:suzuki_time} by a sequence of exponentials in the form:
\begin{eqnarray}
\label{eq:approx_time}
\psi(t_{n+1})
&\approx & \prod_{k=1}^{N_{\rm seq}}\left[e^{-is_{0}^{(k)}\Delta t \mathcal{T}} \prod_{j=1}^{N}e^{-is_{j}^{(k)}\Delta t H_{j}(t_{n})} \right] \psi(t_{n}) \nonumber \\
\end{eqnarray}   
where the coefficients $N_{\rm seq}\in \mathbb{N}$ and $s_{j}^{(k)} \in \mathbb{R}$ are chosen to obtain an approximation with a given order of accuracy. It is then straightforward to conclude that the QLB scheme, shown in Eq. \eqref{eq:QLB_exp} and in Eqs. \eqref{eq:split1} to \eqref{eq:split4}, corresponds 
to a particular decomposition of the Hamiltonian\footnote{The decomposition is such that $H_{1} = -i\alpha_{x}\partial_{x}$, $H_{2} = -i\alpha_{y}\partial_{y}$, $H_{3} = -i\alpha_{z}\partial_{z}$ and $H_{4} = \beta m$.} and to a specific realization of Eq. \eqref{eq:approx_time}.

The conclusion is far reaching; the Majorana representation exposes a
concrete connection between the (discrete) Boltzmann equation and the
Dirac equation in Majorana form.
As a result, the information contained in the
quantum relativistic four-spinor $\psi(t,\mathbf{x})$ can 
be processed on entirely classical terms, i.e
free-streaming along constant directions and local collisions, complemented
with diagonalization steps to keep speed and spin constantly aligned.
Remarkably, the scheme is also viable for prospective quantum 
computer implementations \cite{feynman1982simulating,SLoyd,boghosian1998simulating,Yepez}. 

The QLB has been applied to a variety of quantum wave problems, mostly 
in the non-relativistic context,  \cite{Succi2002317,PhysRevE.75.066704,PhysRevE.77.066708}.
Here we present a new application to an important non-linear relativistic
problem, namely the Dirac equation augmented with Nambu-Jona-Lasinio 
dynamic symmetry breaking terms.

{\it The NJL-Dirac equation}

The Nambu-Jona-Lasinio (NJL) model was prompted out by a profound analogy between the
Bardeen-Cooper-Schrieffer theory of superconductivity and chiral
symmetry breaking in relativistic quantum field theories \cite{NJL1,NJL2}
and it has served ever since as a model paradigm to study 
symmetry-breaking phenomena in both fields. 

The NJL Lagrangian reads \cite{NJL1}
\begin{equation}
\label{eq:NJL_lag}
\mathcal{L}_{NJL} = \bar \psi (i \gamma^{\mu} \partial_{\mu}  -m) \psi 
+ \frac{g}{2}[ (\bar \psi \psi)^2 - (\bar \psi \gamma^5 \psi)^2].  
\end{equation}
This corresponds to the free-particle Dirac Lagrangian, plus an 
interaction term, driven by the coupling parameter $g$. This coupling term reflects four-fermion interactions, in direct
analogy with the BCS theory of superconductivity.
By imposing the chiral symmetry, the NJL lagrangian 
should not present any explicit bare mass term, so we set $m=0$. 
However, the NJL dynamics leads to the formation of
a chiral condensate, corresponding to an effective mass term and a 
spontaneous symmetry breaking of the chiral symmetry.
Much of the current interest in the NJL model is motivated by the fact that
it serves as a phenomenological model of quantum chromodynamics (for a full account see \cite{RevModPhys.64.649}).
 
The associated equation of motion reads (see Appendix)
\begin{eqnarray}
\label{eq:dirac_NJL}
 (  \partial_t + \alpha_a \partial_a + i m \beta ) \psi = ig\beta [ (\psi^{\dagger} \beta \psi)
   - ( \psi^{\dag} \beta
\gamma^5 \psi)  \gamma^5 ]  \psi.\nonumber \\
\end{eqnarray}
A solution of this equation is required for the quantum study of this model, in the mean-field approximation.

The space-time discretization of NJL-Dirac can be cast in the standard
QLB format by adding the non-linear term into the collision step as in the case of the electromagnetic field, by replacing $C \rightarrow C_{\rm NJL}$. The collision step becomes
\begin{eqnarray}
\label{eq:sol_ex_4}
 \psi^{(c)}(t_{n+1}) &=&  C_{\rm NJL} \psi^{(c)}(t_{n}), \nonumber    \\ 
&=&T \exp \left[  \int_{t_{n}}^{t_{n+1}} dtM_{\rm NJL}(t) \right] \psi^{(c)}(t_{n}), \\
M_{\rm NJL}(t) &\equiv &  -i\beta (m - g \rho_{S}(t)) - g\rho_{A}(t) \Sigma ,
\end{eqnarray}
where $\rho_S \equiv \psi^{\dag} \beta \psi$ and $\rho_A \equiv i \psi^{\dag} \beta
\gamma^5 \psi$ depend on time, hence the time-ordering operator, and $\Sigma \equiv \beta \gamma^{5}$. The time-ordering can be approximated by using Eqs. \eqref{eq:suzuki_time} and \eqref{eq:approx_time}: the ensuing ordinary exponential can be converted exactly to a $4\times 4$ unitary matrix $C_{\rm NJL}$. A similar treatment of the nonlinear term, albeit using spectral methods, can be found in \cite{Xu2013131}. 

{\it Numerical application} 

As an application of the QLB scheme, we simulate the emergence
of a dynamic fermion mass as a result of the spontaneous
breaking of the chiral symmetry of the NJL equation. 

\begin{figure}[htbp]
\subfloat[]{\includegraphics[width=0.25\textwidth]{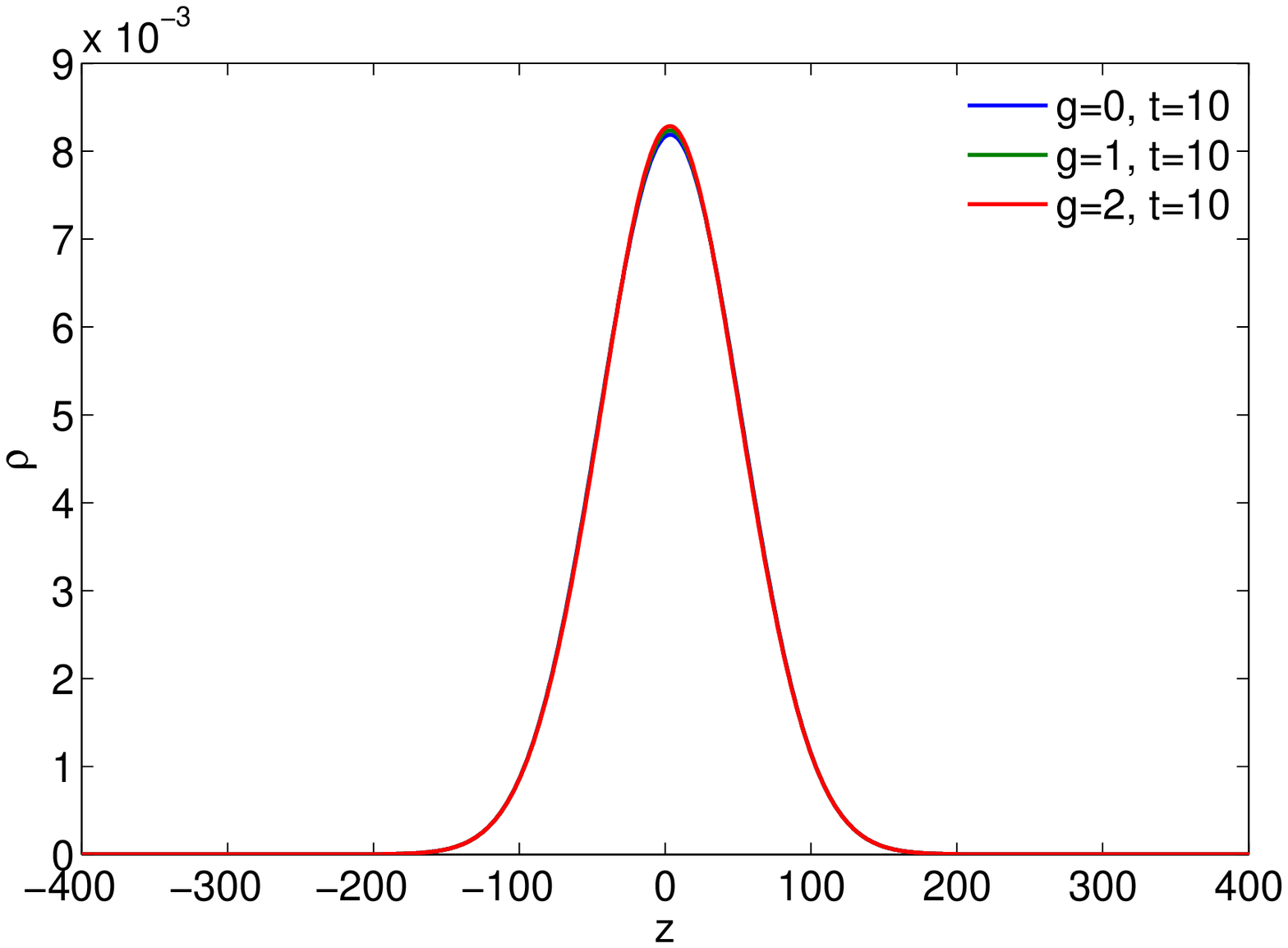}}
\subfloat[]{\includegraphics[width=0.25\textwidth]{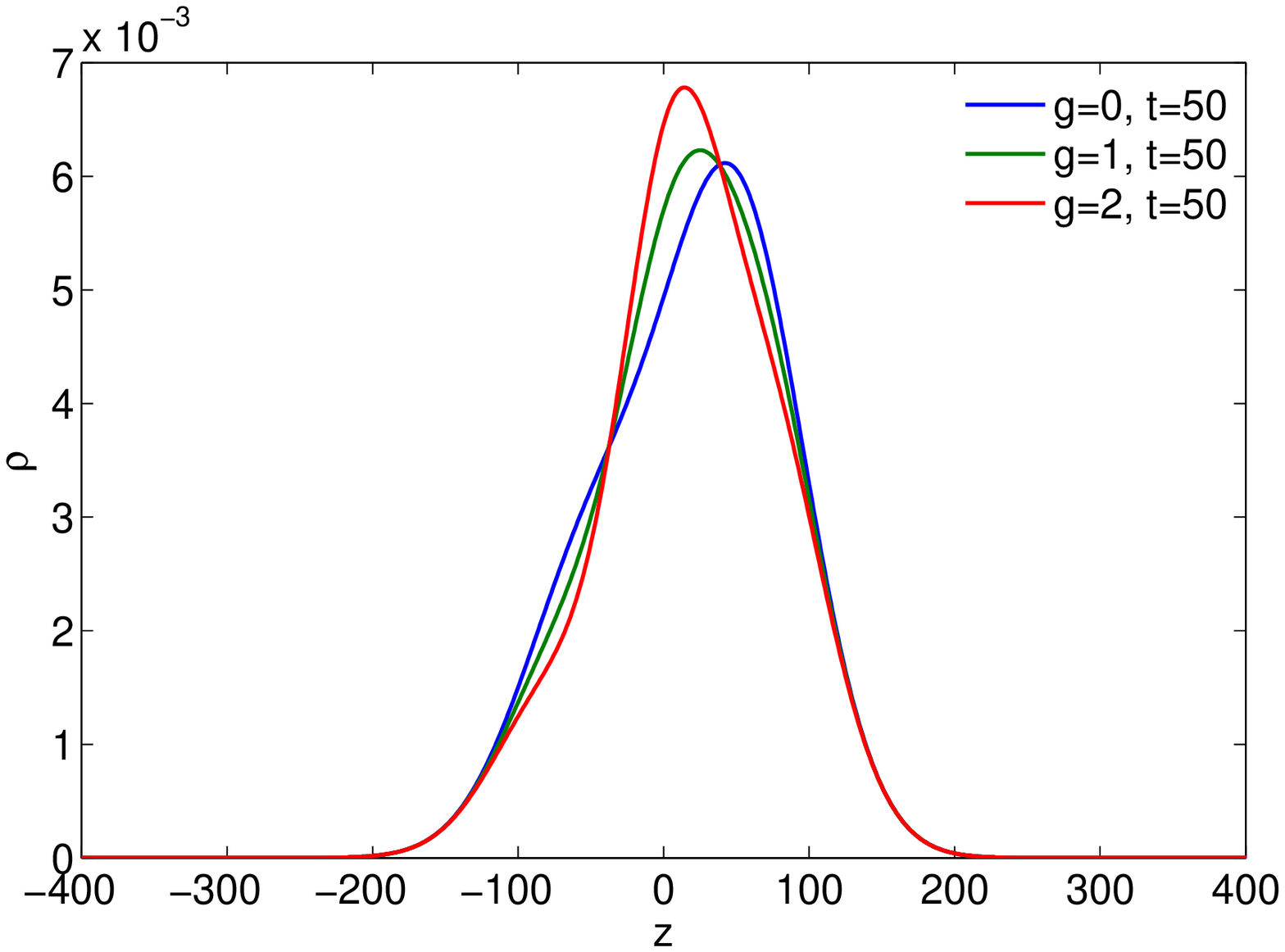}}\\
\subfloat[]{\includegraphics[width=0.25\textwidth]{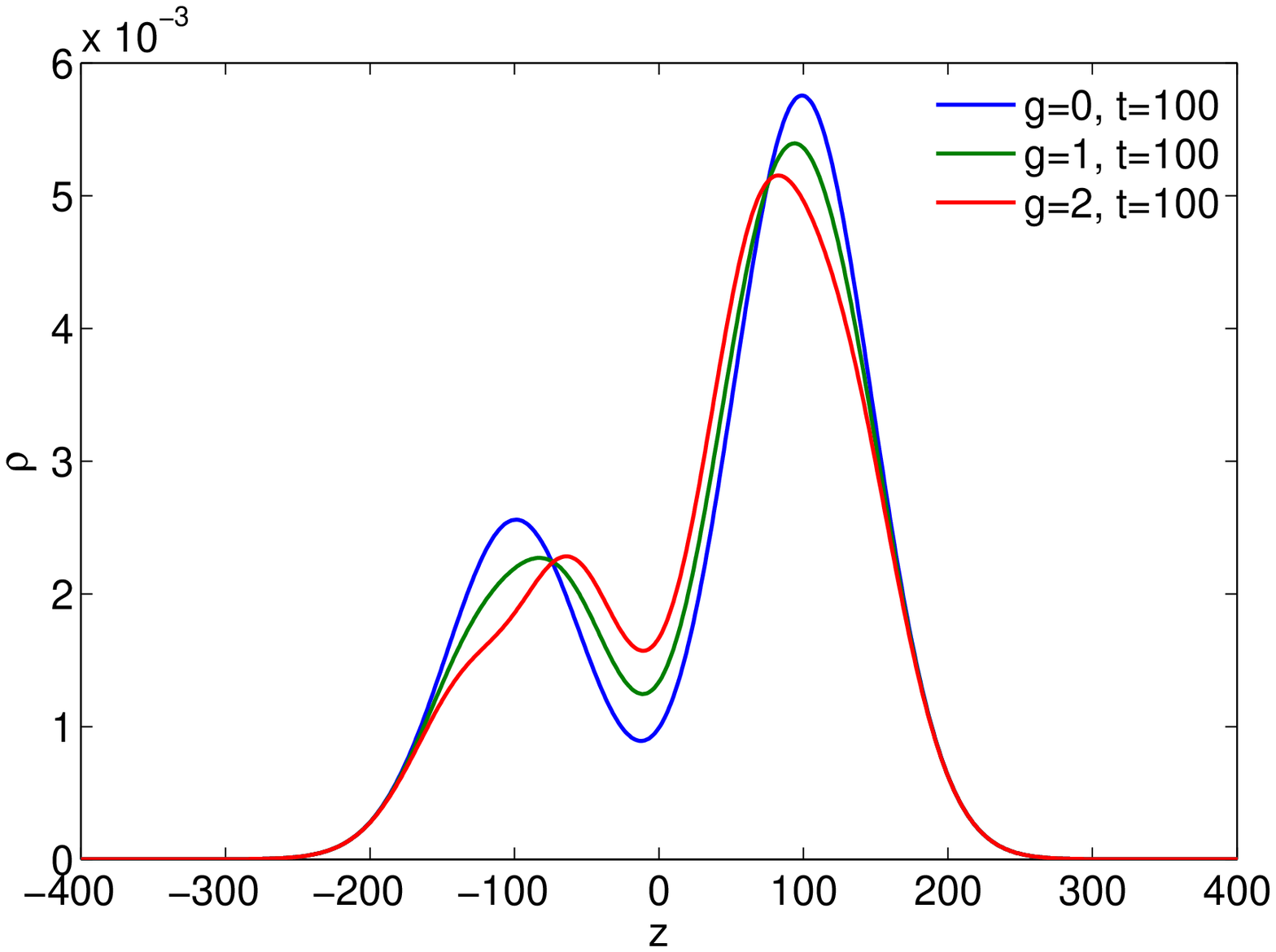}}
\subfloat[]{\includegraphics[width=0.25\textwidth]{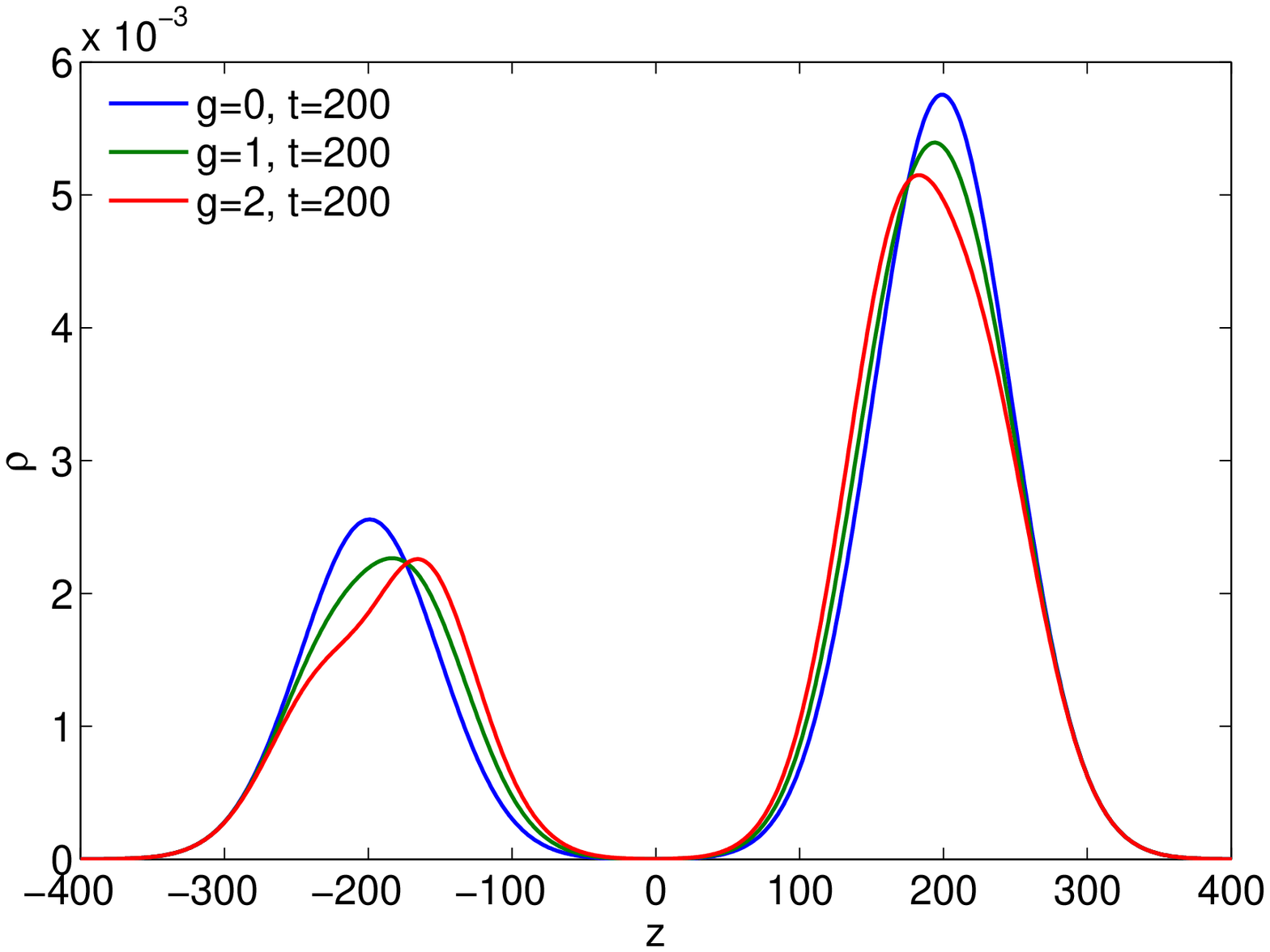}}
\caption{$\rho = |\psi|^2$ at
times (a) $t=10$, (b) $t=50$, (c) $t=100$ and (d) $t=200$ for $g=0$, $1$ and
$2$. The figure shows the separation of the left and right moving wavepackets in the course
of the evolution. The non-interacting case shows no deformation of the Gaussian profile,
as expected, while the interacting case leads to a slow-down and deformation of both
wavepackets. 
}
\label{fig:res_1D}
\end{figure}

For this purpose, let us consider an initial condition given by the following Gaussian 
minimum-uncertainty wave packet
\begin{eqnarray} \label{eq:1D_IC}
\psi(t=0,z) = S_{y} 
\begin{bmatrix}
-C_{u}e^{ikz} + C_{d}e^{-ikz} \\
C_{u}e^{ikz} - C_{d}e^{-ikz}\\
C_{u}e^{ikz} +C_{d}e^{-ikz} \\
C_{u}e^{ikz} +C_{d}e^{-ikz}
\end{bmatrix}
\cfrac{e^{-\frac{z^{2}}{4\sigma^{2}}}}{(2\pi \sigma^{2})^{\frac{1}{4}}}
\end{eqnarray} 
centered about $z=0$, with initial width $\sigma$.
Let $\omega=k$ be the initial energy of the wave packet. 
The coefficients $C_u$ and $C_d$ obey the condition
$2 C_u^2 + 2 C_d^2 = 1 $,
so that $\psi^{\dag}\psi =
|G_0|^2$.
Moreover, an asymmetry can be
set by tuning the ratio $C_u/C_d \equiv \alpha \neq 1$.\\
We analyze our numerical results for the case of $m=0$, which ensures that
the axial current is conserved by the free part of the equation, as a function of the coupling coefficient $g$.
For this test, the following parameter setting is used: $k=0.006$, $\sigma = 48$,
$C_{u}=1.177$ and $C_{d}=0.784$. Numerical results for $\rho(z) = |\psi|^{2}$ 
 at times $t=10$, $50$, $100$ and $200$, for the case $g=0$, $1$ and
 $2$ are shown in Fig. \ref{fig:res_1D}. 

This calculation requires 
200 time-steps (for a mesh size of 1024 lattice sites) and 
about 0.01 CPU seconds on a standard PC. 
This amounts to a processing speed of about 20 MLUPS (Million lattice 
updates per second), which is in line with the performance 
of Lattice Boltzmann schemes for classical fluids. Since the latter is known to be 
very competitive, the same conclusion is likely to
hold for the quantum case. A final statement in this direction
must be left to detailed head-on comparison between QLB and state-of-the
art numerical methods for the Dirac and Schr\"{o}dinger equations.
 
From Fig. \ref{fig:res_1D}, a symmetry breaking between the left and right moving wavepackets 
is clearly seen at increasing values of $g$.
The generation of a dynamic mass is expected to reflect into a slowing-down
of the group velocity of the wavepackets, according to
$\frac{v_{\rm group}}{c} = \frac{k}{\sqrt{k^2+m'^2}} <  1 $
where $m'=-g \rho_S$ is the dynamic mass of zero-rest mass particles.
Indeed, since the initial condition is symmetric with respect to  $1 \leftrightarrow 2$
exchange, the quantity $\rho_A$ is initially zero, and remains such all along the simulation.\\ The results can be checked against
the analytic solution to Eq. \eqref{eq:dirac_NJL} in 1-D and for the case of small $g$ \cite{PRS13}, which gives  $v_{\rm mean}/c\simeq 1-0.04 g +{\cal O}(g^2)$ at early times.
It can be checked (not shown for space limitations) that this is consistent with the 
numerical results in Fig. \ref{fig:res_1D}.\\
The same phenomenon can be simulated in two dimensions, and the details shall be
presented in a future and lenghtier publication. 

Extending the above work to the case of quantum many-body systems and 
non-linear multidimensional quantum field theory \cite{1751-8121-40-26-F07}, represents an outstanding 
challenge for future research in the field.



{\it Appendix: NJL-Dirac equation using Pauli representation}

From the NJL Lagrangian of Eq. \eqref{eq:NJL_lag}, the associated equation of motion Eq. \eqref{eq:dirac_NJL} is derived as follows.
Variation of Eq. \eqref{eq:NJL_lag} against $\bar \psi$ delivers
\begin{equation}
\label{eq:DiracNJL}
(i \gamma^{\mu} \partial_{\mu} -m ) \psi + g \left[ (\bar \psi \psi) \psi +
(\bar \psi \gamma^5 \psi) \gamma^5 \psi \right] = 0,
\end{equation}
where $\gamma^5 \equiv i \gamma^0 \gamma^1 \gamma^2 \gamma^3$ and $\bar \psi =
\psi^{\dag} \gamma^0$. \\
The actual definition of the gamma matrices
depends on the specific chosen representation.
By using
Pauli-Dirac representation, $\gamma^i$ matrices are
defined as follows \cite{DIRAC}:
\begin{equation} \label{eq:gammaPauliDirac}
\begin{split}
\gamma^0 = \beta, \quad
\gamma^i = \beta \alpha^i, \quad \text{with} \quad i=1,\dots,3,
\end{split}
\end{equation}
where $\beta$ and $\alpha^i$ are the standard Dirac
matrices. \\ Inserting these definitions into Eq. \eqref{eq:DiracNJL}, yields
\begin{eqnarray}
 ( \beta \partial_t + \beta \alpha_a \partial_a + i m ) \psi = ig [ (\psi^{\dagger} \beta \psi)
   - ( \psi^{\dag} \beta
\gamma^5 \psi)  \gamma^5 ]  \psi. \nonumber \\
\end{eqnarray}

%

\bibliographystyle{apsrev}

\bibliography{bibliography}

\end{document}